# Research on Human Dynamics of Information Release of WeChat Users

Juliang Zhang[1,a], Shengtai Zhang[1,b], Fan Duo[2,c], Feifei Wang[1,d]

[1] School of economics and management, Beijing University of Posts and Telecommunications, Beijing 100876, China;

[2] School of information and communication engineering, Beijing University of Posts and Telecommunications, Beijing 100876, China.

[a] 944045083@qq.com, [b] sttzz@163.com, [c] fanduo@bupt.edu.cn, [d] wff0102@126.com

**Abstract:** The information release behavior of WeChat users is influenced by many factors, and studying the rules of the behavior of users in WeChat can provide theoretical help for the dynamic research of mobile social network users. By crawling WeChat moments information of nine users within 5 years, we used the human behavioral dynamics system to analyze users' behavior. The results show that the information distribution behavior of WeChat users is consistent with the power-law distribution for a certain period of time. Meanwhile, there is an anti-memory characteristic in information release behavior of WeChat users, which is significantly different from other user behavior patterns in online social networks. The results of the study provide theoretical support for the further study of information release behavior of Wechat users.

## 1. Introduction

With the development of the Internet, people's lifestyles have undergone tremendous changes. Every day people may do a lot of repeated things, but the time and duration of these things is not simply repeated. The time of human behavior is not statistically distributed in the Poisson distribution [1]. In the mobile social network, the role of users in the network is more and more important, and the user is not only the information communicator, but also has become the information producer. WeChat is a social network based on mobile networks. Users can publish text, pictures, videos and links at any time and share anything with others. Traditional networks are content-centric, while mobile social networks are user-centric [2]. With the increasing influence of WeChat, mobile social network has become the main force to promote the social interaction of users. It is very important to study the information dissemination of WeChat users and to find the social law and promote the development of social networks.

The time interval of human activities on the Internet is neither always independent to each other, nor has the same distribution, it follows the power-law distribution with the fat tail[3]. In 2005, Barabási analyzed the data of 3188 users in the university, and found that the individual's response time and interval time were subject to the power-law distribution [4]. Zhou Tao's analysis of the characteristics of time and space in human behavior suggests that although the time interval distribution of human behavior in different systems is different, there are more fat tails than the distribution predicted by Poisson distribution [5]. WeChat is the new media under current mobile social network, which information dissemination intention and behavior has been in-depth study [6-10]. so its users' information release behavior is a new challenge.

## 2. Related Work

With the continuous development and improvement of Internet technology and large data technology, scholars obtain large-scale record of social network behavior data through the network platform open API（Application Program Interface）or network data crawler algorithm. This data provides basis for the rule of social network users and its impact on information dissemination behavior.

In the research of social network user behavior, such as web browsing [11], Email [12], instant messaging [13], microblogging [14], Wikipedia [15], Github [16] and so on, domestic and foreign scholars have made a series of valuable research results. Yan et al studied the release data of 14186 microblogging users from microblogging group. He analyzed it and concluded that the time interval of user's information release complies with power-law distribution whose exponent is 1.3 [17]. Yan Qiang et al collected the number of fans, concerns and microblogging, calculated the time interval of the users' two constant behaviors, found that the time interval obeys the power law distribution for two consecutive microblogging behaviors of a single user, power index of 1.7 [18]. Leskovec collected data from instant messaging tool, MSN, including about 30 billion conversations between 240 million users. The results show that the time interval among users at the group level is subject to the power distribution of the index of 1.5 [19]. Radicchio concluded that the index is 1.2 based on the analysis of the user's landing time interval distribution in Wikipedia [20]. Chen et al study the time interval of QQ user behavior through the record data from five volunteers in the QQ about 1 year; found that time interval obeys the power distribution between 2.0 and 2.5 at the individual level [21].

Through the empirical study of a large number of online social network datasets, it is found that social network users have similar statistical laws, that is, the behavior interval is heterogeneous and paroxysmal in most cases, it can be illustrated by $P(\tau) \sim \tau^{-\alpha}$ ($\tau$ represents the time interval of two consecutive acts).As a mobile social network under the emerging media, whether it has obvious heavy-tail characteristics of the information publishing time interval by WeChat users? Or what is the law of paroxysmal and memory on publishing behavior? These are the two questions we attempt to solve. However, due to the semi-closed and privacy of WeChat, the data is not easy to obtain, so it increased difficulties to our study.
Therefore, we obtained data from WeChat moments, and researched the rule of information released behavior from the perspective of time interval after several experiments, according to the characteristics of WeChat release. The paper is organized as follows: In the third part, we introduce our data sources. In the fourth part, we carry out the analysis of human behavior dynamics, including the distribution of time interval and analysis of paroxysmal and memory characteristic in 5 hours. The fifth part, we conclude our research contributions and practical prospect to this work.

## 3. Datasets description

The data is collected from the WeChat moments, as it has the closed characteristics, and the moments cannot be displayed on the webpage, also there is no open API interface, which makes a new challenge for data acquisition and human dynamics analysis. Making use of the feature that WeChat moments and QQ space can be synchronized to release information, we obtained the data from 9 volunteers from QQ space using crawling software. The synchronization rate of WeChat friends circle and QQ space is up to 95%.

As shown in Table 1, the time span of data differs because it is also different when each user starts to use WeChat. The number of messages is not alike as each user has a different frequency of posting messages on the moments. The dataset includes the user information, the release time and other statistical information. The time span is about three years, and larger amount of data is sufficient to reflect the real statistical laws.

Table 1. Summary information of moments

| Users | The number of messages | Time span (month) |
|---|---|---|
| A | 417 | 24 |
| B | 388 | 49 |
| C | 1062 | 45 |
| D | 488 | 27 |
| E | 426 | 63 |
| F | 2807 | 35 |
| G | 474 | 23 |
| H | 1199 | 30 |
| I | 1641 | 57 |

**4. Experiment and Analysis**
**4.1 The distribution of user's time interval**
**4.1.1 Distribution of information release behavior time interval**

The time statistic characteristics of human behavior is not uniform distribution, because of long time silence and the outbreak of high frequency in a short time, The time interval distribution of successive behaviors has obvious fat tail characteristics, This paper fit the power law by MLE (Maximum Likelihood Estimation) estimation

CDF (Cumulative Distribution Function) method which was proposed by Ethan [22]. It can be better fitted with a power law distribution function, ie:

$$P(\tau) \sim \tau^{-\alpha} \tag{1}$$

In the case where the original distribution satisfies (1), the cumulative distribution satisfies:

$$P(X \gg x) = C \int_x^\infty x^{-\alpha} dx = \frac{C}{\alpha-1} x^{-(\alpha-1)} \tag{2}$$

According to the time data of user's information release, we study the time interval of two consecutive acts. Figure 1 shows the cumulative distribution of the time interval of the information releasing behavior by WeChat users. The horizontal axis represents the time interval and the vertical axis represents the cumulative probability density.

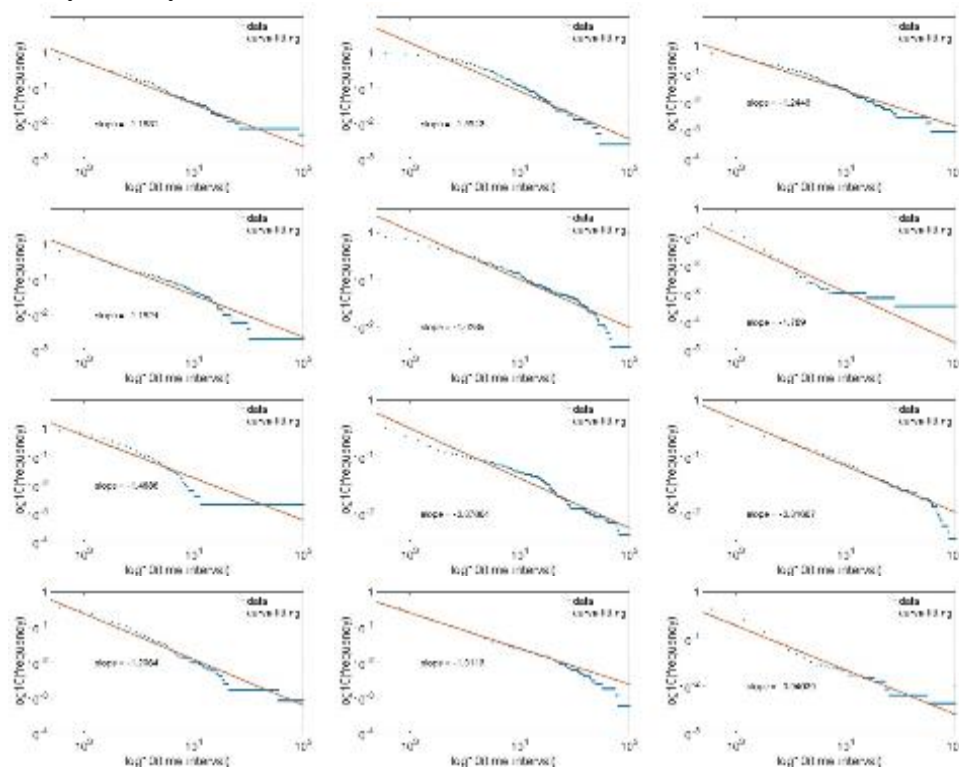

Figure 1. The time interval distribution of information release behavior by WeChat user

As shown in Figure 1, the graph has a large tail and there are obvious "fat tail" features, It is showed that the fitted power exponents are between 1 and 2 which of the information releasing behavior from nine users of WeChat. And that is similar to the time interval distribution power exponent of the behavior in the web browsing[23], online movie reviews[24] and social network[25], which is between 1 and 3, all these indicate that the releasing behavior has power-law features. It also shows that in the time interval distribution of majority users, a small amount of time interval is very long, that is, a small number of information interval, and large number of information interval is small. And most number of time intervals is short, which is that most of the release behavior will occur intensively in a short time, and then there will be a very long gap.

### 4.1.2 Paroxysmal

When the time interval distribution obeys the power function law, sometimes many events will occur intensively in a short time, and then there will be a long gap period, this situation is called paroxysmal.

$$B = \frac{\sigma_\tau - m_\tau}{\sigma_\tau + m_\tau} \tag{3}$$

in (3), where $\sigma_\tau$ and $m_\tau$ represent the standard deviation and mean of the distribution $p(\tau)$, respectively, and the value of $B$ is between -1 and 1. For the Poisson distribution, the mean and standard deviations are equal, so the paroxysmal is exactly zero; for a severe fat tail distribution, the standard deviation is much greater than the mean, $B$ is close to 1. For the periodic rule signal, the time interval distribution is actually a $\delta$ Function, the standard deviation is 0, $B$ value is -1.

Table 2. Paroxysmal of 9 users

| users | A | B | C | D | E | F | H | I | J |
|---|---|---|---|---|---|---|---|---|---|
| $B$ | 0.9033 | 0.8953 | 0.9359 | 0.9097 | 0.9072 | 0.9608 | 0.9091 | 0.9412 | 0.9469 |

As shown in Table 2, $B$ is the value of nine users and it is close to 1. There is a serious fat tail distribution, which indicates that when WeChat users release information, the majority released in a shorter time interval, and then there is a long gap. This also reflects the fragmentation characteristics of WeChat as an instant messaging tool. When the user is in the spare time, to publish information in the fragmented time using WeChat, the publishing behavior will occur intensively in a short period of time, and then there will be a very long gap.

### 4.1.3 Memory

The time series of events can also be characterized by memory. That is, the time series of a behavior is considered to be memorial if the long time interval is followed by a longer time interval and the short time interval is followed by a shorter time interval. If all the time intervals are preceded by a sequence (Note: it is not a sequence of behavioral occurrences but a sequence of two successive behavior intervals), the sequence has a total of $n_\tau$ elements (which means that there is a total behavior of $n_\tau+1$) The first $n_\tau-1$ elements constitute the sequence 1, and the last $n_\tau-1$ constitute the sequence 2, As shown in the formula, the Pearson association of the two sequences can be used to measure the memory of the sequence.

$$M = \frac{1}{n_\tau - 1} \sum_{i=1}^{n_\tau - 1} \frac{(\tau_i - m_1)(\tau_{i+1} - m_2)}{\sigma_1 \sigma_2} \tag{4}$$

In the formula, $m_1$ and $m_2$ are the mean of sequence 1 and sequence 2, respectively; $\sigma_1$ and $\sigma_2$ are the standard deviations of sequence 1 and sequence 2, respectively. Obviously, the value of M is in the range of -1 to 1: M> 0 means memory effect, M <0 means anti-memory effect.

Table 3. Memory of 9 users

| users | A | B | C | D | E | F | H | I | J |
|---|---|---|---|---|---|---|---|---|---|
| M | -0.0081 | -0.0044 | -0.0127 | -0.0225 | -0.0057 | -0.0105 | 0.007 | -0.0103 | -0.0033 |

As shown in Table 3, only one of the 9 users showed weak memory, while the other eight users showed weak anti-memory, and the short (long) time interval was usually accompanied by a long (short) time interval. This is consistent with paroxysmal results. It indicates that WeChat release behavior is different from other social networks, based on the semi-closed strong relationship network, the users will be more casual when they release messages.

**4.2 Time interval distribution at different intervals**

9 users' power exponents are between 1 and 2, but in the different scales such as different minutes, hours and days the exponents are not alike, we select three representative users based on the different time span: A, H, I, their time span are 2 years, 3 years, 4 years.

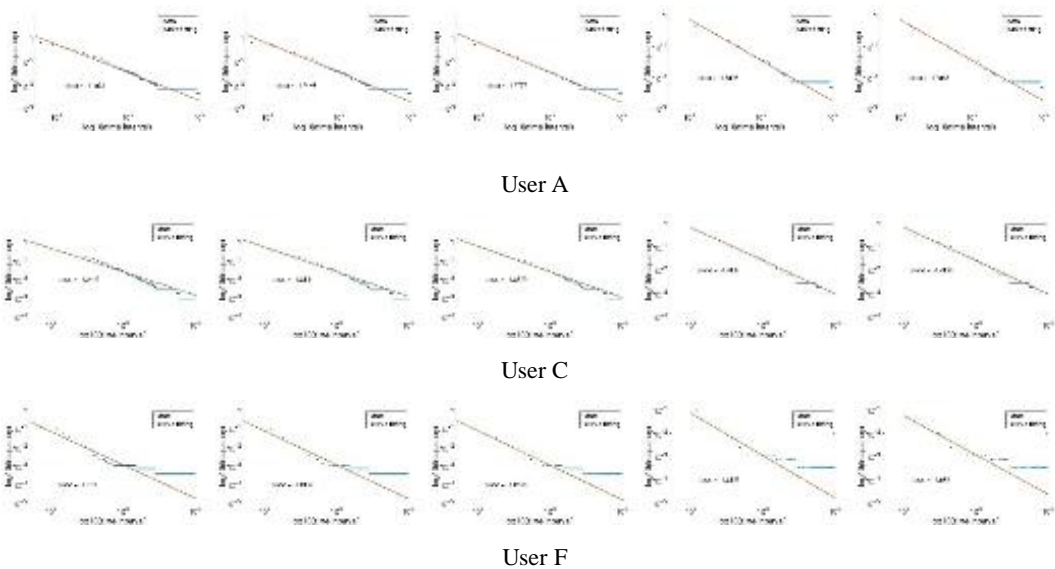

User A

User C

User F

Figure 2. Time interval distribution (1-5h)

As shown in Fig. 2, the distribution of three users is subject to the power exponent distribution at different time intervals of 1h, 2h, 3h, 4h, and 5h by adjusting the time interval of three users. The power exponents of three users are increasing with the increase of time interval, which indicates its semi-closed characteristics, so that the privacy of friend circle is guaranteed. At the same time, WeChat moments is a kind of strong network, the user's friends are mostly having the same characteristic with the user himself, so the information published in the same circle will have very small differences.

**5. Discussion and Conclusion**

In this paper, through the dynamic study of the time interval of user information release by WeChat users, we find that the information release behavior confirms to the time characteristics of human dynamics. It indicates that most users' publishing behavior will occur intensively in a relatively short time. Second, the release behavior shows a weak anti-memory property, and the publishing behavior usually follows a shorter time interval after a long time interval, and a short time interval is followed by a longer time interval. Third, the distribution of WeChat users is different from other online networks behaviors, WeChat moments is a network of strong relations, its behavior is basically similar and the internal differences are small.

The sample taken was nine volunteers, there are some limitations of the number of samples, and we didn't take into account of the behavioral patterns of different user characteristics. In addition, this study focuses on the user release time interval rule without considering other factors, such as the number of reviews, reading and so on. In addition to expanding the number of samples, the future work should also consider the behavior of different groups, such as gender, age, occupation, education and so on. At the same time, we can consider extending the information release behavior to the study of the comment and "thumbs up", and make further study on the information distribution law of the WeChat users.

## Acknowledgments

This work was sponsored by National Natural Science Foundation of China (NSFC) under grant number 71571022.